\documentclass[aps,pra,twocolumn,nofootinbib,10pt]{revtex4-1}

\usepackage{amsmath,amssymb,bm}
\usepackage{graphics}

\newcommand{\im}{i}
\newcommand{\dm}{d}
\newcommand{\tr}{\text{tr}}
\newcommand{\eqtri}{\triangleq}
\newcommand{\avg}[1]{\langle{#1}\rangle}
\newcommand{\Dcirc}{{\cal D}^{\circ}}

\begin{document}

\title{Evolution of moments in atmospheric scintillation}
\author{Filippus S. \surname{Roux}}
\email{rouxf@ukzn.ac.za}
\affiliation{University of Kwazulu-Natal, Private Bag X54001, Durban 4000, South Africa}

\begin{abstract}
Evolution equations for the moments of a photonic quantum state propagating through atmospheric turbulence are derived. These evolution equations are obtain from an evolution equation for the characteristic functional of the state, incorporating all spatiotemporal degrees of freedom. The measured quantities, such as the intensity or photon distribution, of the evolving state can be expressed in terms of such moments without having to know the exact final state. The case of an initial coherent state is considered as an example.
\end{abstract}

\maketitle

\section{\label{intro}Introduction}

The evolution of a photonic quantum state propagating through a turbulent atmosphere is a challenging physical process to analyze, despite several theoretical studies \cite{paterson,sr,qturb4,qturb3,ipe,turbsim,leonhard,semenov,notrunc,ipfe} and physical demonstrations \cite{malik,vallone,krenn2,qkdturb}. An understanding of such scenarios is relevant for free-space quantum communication systems, implementating quantum cryptography and continuous variable teleportation \cite{cvteleport,wig1,semenov,fstelez,heim,cvcomrev,advtelport,cvqkd,derkach,cvtelml}. Part of this challenge is due to the multiphoton  nature of the photonic quantum states that are used in some of these scenarios.

Despite the quantum nature of such an analysis, it can be argued that the same procedure can be used for purely classical applications. In fact, while the basic phase modulation process that is at work in the atmospheric scintillation of optical fields is a classical process, its effect on bright optical fields can lead to quantum effects that may affect the way such fields are scintillated. Regarded as quantum states, such bright optical fields are multiphoton states. It is known that multiphoton states can lead to quantum interference effects even when such states propagate through classical optical components such as beamsplitters, as demonstrated by the Hong-Ou-Mandel effect \cite{hongoumandel}. It is thus argued that the random index fluctuations in a turbulent atmosphere can also lead to multiphoton interference effects. In this way, classical scintillation may benefit from an analysis that treats the classical optical field as a multiphoton quantum state and incorporates the scintillation process as a phase modulation on a photon-per-photon basis.

Here, we use a functional phase space approach in which an evolution equation for the characteristic functional of the multiphoton state \cite{ipfe} is used to obtain evolution equations for the moments of these states. These evolution equations can be solved in an iterative process, based on the initial moments of the input state. The moments provide sufficient information to express measured quantities of the state after having propagated through the scintillating medium.

In this work, we first review the derivation of the evolution equation for the characteristic functional, starting from the derivation of the evolution equation for the Wigner functional of the state. Then we derive evolution equations for the moments. Finally we apply these equations to the case of an initial coherent state. By working through the first few orders, we are able to discern a pattern that can be used to express the photon-number distribution to all orders in the scintillation kernel.

\section{Derivation of the evolution equation}

Given the paraxial propagation equation for classical light in a turbulent atmosphere,
\begin{equation}
\partial_z g(\mathbf{x},z) = \frac{\im}{2k} \nabla_{\perp}^2 g(\mathbf{x},z) + \im k \tilde{n}(\mathbf{x},z) g(\mathbf{x},z) ,
\end{equation}
in which $g(\mathbf{x},z)$ is the slow-varying scalar electromagnetic phasor field as a function of the two-dimensional transverse coordinate vector $\mathbf{x}$ and the propagation distance $z$, $k=2\pi/\lambda$ is the wavenumber, $\nabla_{\perp}^2=\partial_x^2+\partial_y^2$, and $\tilde{n}(\mathbf{x},z)$ is the fluctuation in the refractive index of the atmosphere (the refractive index is $n=1+\tilde{n}$), we apply a transverse Fourier transform to obtain the equivalent equation for the angular spectrum
\begin{equation}
\partial_z G(\mathbf{k},z) = \im \int P_{\text{c}}(\mathbf{k},\mathbf{k}') G(\mathbf{k}',z)\ \frac{\dm^2k'}{(2\pi)^2}
\eqtri \im P_{\text{c}}\diamond G .
\label{eom}
\end{equation}
Here, $\mathbf{k}$ is the two-dimensional transverse wavevector,
\begin{equation}
G(\mathbf{k},z) \eqtri \int g(\mathbf{x},z) \exp\left(-\im \mathbf{k}\cdot\mathbf{x}\right)\ \dm^2x
\end{equation}
is the evolving angular spectrum, and
\begin{equation}
P_{\text{c}}(\mathbf{k},\mathbf{k}',z) \eqtri k N(\mathbf{k}-\mathbf{k}',z)
- \mathbf{1}(\mathbf{k},\mathbf{k}') \frac{|\mathbf{k}'|^2}{2k}
\label{defmc}
\end{equation}
is the propagation kernel for the scintillating medium in a {\em co-moving reference frame}, where
\begin{equation}
\mathbf{1}(\mathbf{k},\mathbf{k}') \eqtri (2\pi)^2 \delta(\mathbf{k}-\mathbf{k}') ,
\end{equation}
is the identity kernel, and
\begin{equation}
N(\mathbf{k},z) \eqtri \int \tilde{n}(\mathbf{x},z) \exp\left(-\im \mathbf{k}\cdot\mathbf{x}\right)\ \dm^2x .
\end{equation}
In Eq.~(\ref{eom}), we introduce a simplified notation to represent the integration over a shared transverse wavevector by a $\diamond$-contraction.

The equation can also be expressed in a {\em fixed reference frame} by replacing
\begin{equation}
G(\mathbf{k},z) \rightarrow \exp\left(\frac{\im z|\mathbf{k}|^2}{2k}\right) G_{\text{f}}(\mathbf{k},z) .
\end{equation}
It removes the free-propagation term from $P_{\text{c}}$ and modifies the remaining part of the propagation kernel, so that $\partial_z G_{\text{f}}(\mathbf{k},z)= \im P_{\text{f}}\diamond G_{\text{f}}$, where
\begin{equation}
P_{\text{f}}(\mathbf{k},\mathbf{k}') \eqtri k N(\mathbf{k}-\mathbf{k}',z)
\exp\left[\frac{\im z}{2k}\left(|\mathbf{k}|^2-|\mathbf{k}'|^2\right)\right] .
\label{defmf}
\end{equation}
Altough $G_{\text{f}}(\mathbf{k},z)$ evolves during the propagation process, it is referenced to the input plane. Therefore, the final expression requires a final free-space propagation from the input to the output plane.

The benefit of the classical evolution of the angular spectrum is that it provides us with the expression of the propagation kernel for the atmorpheric scintillation process, which can be used in comparison with the spatial equivalent of the Heisenberg equation to identify the propagation operator in the quantum context that serves as the generator for spatial evolution. This propagation operator can then be used in the spatial equivalent of the von Neumann equation for the photonic quantum state propagating through a scintillating atmosphere. We convert this von Neumann equation into its functional phase space equivalent to obtain an evolution equation for the Wigner functional of the state, using the same procedure that was used to derive an evolution equation for parametric down-conversion \cite{nosemi}.

\subsection{Wigner functionals}

The generic form of an evolution equation in terms of Wigner functionals is
\begin{equation}
-\im \hbar\frac{\dm}{\dm z} W_{\hat{\rho}} = W_{\hat{P}}\star W_{\hat{\rho}}-W_{\hat{\rho}}\star W_{\hat{P}} ,
\end{equation}
where $W_{\hat{\rho}}$ and $W_{\hat{P}}$ are the Wigner functionals for the unknown state and the propagation operator, respectively, and $\star$ represents the Moyal star product. It is more convenient to represent the Wigner functional of the propagation operator in terms of a construction process applied to a generating functional. The generating functional is given by
\begin{equation}
\mathcal{W}[\nu^*,\mu] = \exp\left(\nu^*\diamond\alpha+\alpha^*\diamond\mu-\tfrac{1}{2}\nu^*\diamond\mu \right) ,
\end{equation}
where $\alpha$ is the complex phase space field variable, and $\nu^*$ and $\mu$ are auxiliary field variables. The construction process is implemented in terms of functional derivatives with respect to the auxiliary field variables. It reads
\begin{equation}
\mathcal{C} = \hbar\delta_{\mu}\diamond P\diamond\ \delta_{\nu}^* ,
\label{konstrukn}
\end{equation}
where $P$ is either Eq.~(\ref{defmc}) or Eq.~(\ref{defmf}), and
\begin{equation}
\delta_{\mu}(\mathbf{k}) \eqtri \frac{\delta}{\delta\mu(\mathbf{k})} ~~~ \text{and} ~~~
\delta_{\nu}^*(\mathbf{k}) \eqtri \frac{\delta}{\delta\nu^*(\mathbf{k})} .
\end{equation}

The evolution equation then reads
\begin{equation}
-\im \frac{\dm }{\dm z} W_{\hat{\rho}}
= \left.\mathcal{C}\left\{\mathcal{W}\star W_{\hat{\rho}}-W_{\hat{\rho}}\star\mathcal{W}\right\} \right|_{\mu=\nu^*=0} ,
\end{equation}
where we cancelled the factors of $\hbar$. The star products can be evaluated even though the Wigner functional of the state is unknown. They produce
\begin{align}
\begin{split}
\mathcal{W}\star W_{\hat{\rho}}
= &\ \exp\left(\nu^*\diamond\alpha+\alpha^*\diamond\mu-\tfrac{1}{2}\nu^*\diamond\mu \right) \\
& \times W_{\hat{\rho}}\left[\alpha^*+\tfrac{1}{2}\nu^*,\alpha-\tfrac{1}{2}\mu\right] , \\
W_{\hat{\rho}}\star\mathcal{W}
= & \exp\left(\nu^*\diamond\alpha+\alpha^*\diamond\mu-\tfrac{1}{2}\nu^*\diamond\mu \right) \\
& \times W_{\hat{\rho}}\left[\alpha^*-\tfrac{1}{2}\nu^*,\alpha+\tfrac{1}{2}\mu\right] .
\end{split}
\label{sterprods}
\end{align}
When we apply the construction operation to the two star products, it leads to the evolution equation
\begin{equation}
-\im \partial_z W_{\hat{\rho}}
= \alpha^*\diamond P\diamond\left(\delta_{\alpha}^* W_{\hat{\rho}}\right)
- \left(\delta_{\alpha} W_{\hat{\rho}}\right)\diamond P\diamond\alpha .
\label{ipewig1}
\end{equation}
The total derivative becomes a partial derivative because the field variables are independent of $z$.

\subsection{Characteristic functionals}

The characteristic functional is related to the Wigner functional by a symplectic functional Fourier transform. It is a generating functional for the moments of that Wigner functional. By converting the evolution equation for the Wigner functional of the scintillated state in Eq.~(\ref{ipewig1}) into an evolution equation for the characteristic functional of that state, we obtain an evolution equation for the generating functional of the moments of that state (or a generating functional for their evolution equations).

Such an evolution equation for the characteristic functional of a scintillated state has previous been derived \cite{ipfe}, but without considering any further applications. Nor has the potential for deriving equations of moments from it been discussed.

To convert Eq.~(\ref{ipewig1}) into an an evolution equation for the characteristic functional, we replace the Wigner functional $W[\alpha,\alpha^*]$ by its representation in terms of the characteristic functional $\chi[\eta,\eta^*]$, as given by the inverse symplectic functional Fourier transform
\begin{equation}
W[\alpha] = \mathcal{F}^{-1}\{ \chi \} = \int \chi[\eta] \exp(\alpha^*\diamond\eta-\eta^*\diamond\alpha)\ \Dcirc[\eta] .
\end{equation}
Then we apply a symplectic functional Fourier transform to the entire equation, leading to
\begin{equation}
\chi[\eta] = \mathcal{F}\{ W \} = \int W[\alpha] \exp(\eta^*\diamond\alpha-\alpha^*\diamond\eta)\  \Dcirc[\alpha] .
\end{equation}
The moments of the Wigner functional are now obtained from its characteristic functional with the aid of functional derivatives.
\begin{widetext}
In general
\begin{equation}
\mathcal{M}_{m,n} \eqtri \int \alpha^m \alpha^{*n} W[\alpha]\ \Dcirc[\alpha]
= \left. \left(\delta_{\eta}^*\right)^m \left(-\delta_{\eta}\right)^n \chi[\eta] \right|_{\eta=0} .
\end{equation}
The functional derivatives each carry a wavevector dependence that is tranferred to the moment.

It then follows that
\begin{align}
\begin{split}
\mathcal{F}\left\{\alpha(\mathbf{k}_a) \delta_{\alpha}(\mathbf{k}_b) W \right\}
& = \delta_{\eta}^*(\mathbf{k}_a) \left[-\eta^*(\mathbf{k}_b) \chi\right]
= -\mathbf{1}(\mathbf{k}_b,\mathbf{k}_a) \chi - \eta^*(\mathbf{k}_b) \delta_{\eta}^*(\mathbf{k}_a) \chi , \\
\mathcal{F}\left\{\alpha^*(\mathbf{k}_a) \delta_{\alpha}^*(\mathbf{k}_b) W \right\}
& = -\delta_{\eta}(\mathbf{k}_a) \left[\eta(\mathbf{k}_b) \chi\right]
= -\mathbf{1}(\mathbf{k}_b,\mathbf{k}_a) \chi - \eta(\mathbf{k}_b) \delta_{\eta}(\mathbf{k}_a) \chi .
\end{split}
\end{align}
\end{widetext}
The identity kernels lead to traces of the kernel $P$, which cancel between the two terms. The evolution equation in Eq.~(\ref{ipewig1}) thus becomes the first order evolution equation for the characteristic functional given by
\begin{equation}
-\im \partial_z \chi(z) = \eta^*\diamond P\diamond\delta_{\eta}^*\chi-\delta_{\eta}\chi\diamond P\diamond\eta .
\label{ipekar1a}
\end{equation}

\subsection{Second-order equation}

The evolution equations in Eqs.~(\ref{ipewig1}) and (\ref{ipekar1a}) represent a unitary process, tacitly assuming the medium (represented by $N$) is known exactly. Although there are techiques to determine the properties of the medium, allowing its representation as a finite dimensional unitary process \cite{channel}, we consider here the scenario where we only know some statistical properties of the medium. It thus requires an ensemble averaging process.

In such a situation, the characteristic functional (or Wigner functional) of the state becomes a stochastic quantity. As a consequence, one cannot in general separate the characteristic functional of the state from the kernel $P$ in the ensemble average, leading to an intractable situation. To resolve this issue, we perform repeated back substitutions until the evolving characteristic functional of the state is replaced by that of the initial state, which is a known fixed deterministic functional. For this purpose, the evolution equation in Eq.~(\ref{ipekar1a}) is integrated over $z$, leading to
\begin{align}
\chi(z) = & \chi_0 + \im \int_{z_0}^z \eta^*\diamond P(z_1)\diamond\delta_{\eta}^*\chi(z_1) \nonumber \\
 & -\delta_{\eta}\chi(z_1)\diamond P(z_1)\diamond\eta\ \dm z_1 ,
\end{align}
where $\chi_0$ is the characteristic functional of the initial state. Since $\avg{N}=0$, $\avg{P}$ is either zero in the fixed reference frame or equal to the free-space kinetic term for the co-moving frame. As a result, the back substitutions need to be repeated until the second-order terms with two factors of $P$ contain the characteristic functional of the initial state. For definitiveness, we now assume the fixed reference frame for the subsequent expressions.

\begin{widetext}
Repeated back substitutions followed by ensemble averaging then gives
\begin{align}
\partial_z \chi = & -\int \eta^*(\mathbf{k}_1) V_1(\mathbf{k}_1,\mathbf{k}_2,z)\frac{\delta\chi_0}{\delta\eta^*(\mathbf{k}_2)}
+\frac{\delta\chi_0}{\delta\eta(\mathbf{k}_1)} V_1(\mathbf{k}_1,\mathbf{k}_2,z)
\eta(\mathbf{k}_2)\ \frac{\dm^2 k_1}{(2\pi)^2}\ \frac{\dm^2 k_2}{(2\pi)^2} \nonumber \\
& + \int  V_0(\mathbf{k}_1,\mathbf{k}_2,\mathbf{k}_3,\mathbf{k}_4,z) \left[
2\eta^*(\mathbf{k}_1) \frac{\delta^2 \chi_0}{\delta\eta^*(\mathbf{k}_2)\delta\eta(\mathbf{k}_3)}\eta(\mathbf{k}_4)
- \eta^*(\mathbf{k}_1)\eta^*(\mathbf{k}_3) \frac{\delta^2 \chi_0}{\delta\eta^*(\mathbf{k}_2)\delta\eta^*(\mathbf{k}_4)} \right. \nonumber \\
& \left. - \frac{\delta^2 \chi_0}{\delta\eta(\mathbf{k}_1)\delta\eta(\mathbf{k}_3)} \eta(\mathbf{k}_2)\eta(\mathbf{k}_4)
\right]\ \frac{\dm^2 k_1}{(2\pi)^2}\ \frac{\dm^2 k_2}{(2\pi)^2}\
\frac{\dm^2 k_3}{(2\pi)^2}\ \frac{\dm^2 k_4}{(2\pi)^2} ,
\label{karevolm}
\end{align}
where
\begin{align}
\begin{split}
 V_0(\mathbf{k}_1,\mathbf{k}_2,\mathbf{k}_3,\mathbf{k}_4,z)
\eqtri & \int_{z_0}^z \langle P(\mathbf{k}_1,\mathbf{k}_2,z) P(\mathbf{k}_3,\mathbf{k}_4,z_1)\rangle\ \dm z_1 , \\
 V_1(\mathbf{k}_1,\mathbf{k}_2,z)
\eqtri & \int_{z_0}^z \int \langle P(\mathbf{k}_1,\mathbf{k},z) P(\mathbf{k},\mathbf{k}_2,z_1)\rangle\
\frac{\dm^2 k}{(2\pi)^2}\ \dm z_1  \\
\equiv & \int V_0(\mathbf{k}_1,\mathbf{k}_a,\mathbf{k}_b,\mathbf{k}_2,z)
\mathbf{1}(\mathbf{k}_a,\mathbf{k}_b)\ \frac{\dm^2 k_a}{(2\pi)^2}\ \frac{\dm^2 k_b}{(2\pi)^2} .
\end{split}
\label{phidefs}
\end{align}
\end{widetext}
The result is a first-order derivative in $z$, equated to terms with integrals over $z_1$ from $z_0$ to $z$. These integrals are absorbed into the definitions of $V_0$ and $V_1$. After the back-substitution, all terms contain two $P$'s (for the fixed reference frame), one of which is integrated over $z_1$. Since the initial characteristic functional $\chi_0$ at $z_0$ is not a stochastic quantity, it is removed from the ensemble averaging process. Unless the two $P$'s are contracted on each other, the ensemble averaging combines them into one four-point kernel, $V_0$. Where two of the legs of the four-point kernel are contracted on each other, the result is the bilinear kernel $V_1$.

The evolution equation in Eq.~(\ref{karevolm}) can be interpreted in two different ways. When the $z$-integral extends all the way from the initial input plane at $z=0$ (so that $z_0=0$), it represents a non-Markovian process in which the medium for the entire distance from the input plane is required to produce the characteristic functional of the state at $z$.

Alternatively, we can reduce the integration to an infinitesimal range prior to the plane at $z$. The result is an infinitesimally propagation of a given state at $z=z_0$, still regarded as a fixed non-stochatic functional. Then, we employ the Markov approximation removing the $z$-integration in the definitions of $V_0$ and $V_1$. The resulting equation obtained by taking the limit where $\Delta z=z-z_0\rightarrow 0$ is a pure first-order differential equation. We'll use the latter approach.

\subsection{Kernel expressions}

The definition of $P$ in the fixed reference frame in Eq.~(\ref{defmf}), the definitions of $V_0$ and $V_1$ in Eq.~(\ref{phidefs}), and the relationship between the ensemble average of the refractive index fluctuations and the power spectral density (in a statistically homogenous medium) given by
\begin{align}
\Phi_n(\mathbf{k},k_z) \eqtri & \int \avg{\tilde{n}(\mathbf{x}',z)\tilde{n}(\mathbf{x}'+\mathbf{x},z'+z)} \nonumber \\
& \times \exp(-\im\mathbf{k}\cdot\mathbf{x}-k_z z)\ \dm^3 x ,
\end{align}
are used to compute the expressions for the two kernels. Under the Markovian approximation, we replace $\Phi_n(\mathbf{k},k_z) \rightarrow \Phi_n(\mathbf{k},0)$ to alleviate the evaluation of the $k_z$-integral. The expression for $V_0$ becomes
\begin{align}
V_0 = &\ 2\pi^2 k^2 \delta(\mathbf{k}_1-\mathbf{k}_2+\mathbf{k}_3-\mathbf{k}_4) \Phi_n(\mathbf{k}_1-\mathbf{k}_2,0) \nonumber \\
& \times \exp\left[\frac{\im z}{2k}\left(|\mathbf{k}_1|^2-|\mathbf{k}_2|^2+|\mathbf{k}_3|^2-|\mathbf{k}_4|^2\right)\right] .
\label{defphi0}
\end{align}
Note that this four-point kernel is symmetric with respect to $\mathbf{k}_1\leftrightarrow\mathbf{k}_3$ and $\mathbf{k}_2\leftrightarrow\mathbf{k}_4$.

For $V_1$, the contraction produces a significant simplification in the fixed reference frame, leading to
\begin{equation}
V_1(\mathbf{k}_1,\mathbf{k}_2) = \tfrac{1}{2} \Lambda k^2 \mathbf{1}(\mathbf{k}_1,\mathbf{k}_2) ,
\label{defphi1}
\end{equation}
where
\begin{equation}
\Lambda \eqtri \int \Phi_n(\mathbf{k},0)\ \frac{\dm^2 k}{(2\pi)^2} .
\label{deflam}
\end{equation}

\section{Moments}

We can now use Eq.~(\ref{karevolm}) to derive evolution equations for the moments of the Wigner functional of the state. Since the characteristic functional and the Wigner functional can be converted back into each other, they both contain the complete information of the state. The characteristic functional can be expanded as a Taylor series in terms of the auxiliary variables in which the coefficients are the moments of the Wigner functional. Therefore, the moments of the Wigner functional contain all the information of the state. Moreover, all measurements applied to the state can be expressed in terms of these moments.

For example, the intensity or photon-number distribution of a state, which is obtained from being traced with a localized number operator, is given by
\begin{align}
\avg{n} & = \tr\{\hat{\rho}\hat{n}_D\} \nonumber \\
& = \int W_{\hat{\rho}}[\alpha] (\alpha^*\diamond D\diamond\alpha-\tfrac{1}{2}\tr\{D\})\ \Dcirc[\alpha] ,
\end{align}
where $D(\mathbf{k}_1,\mathbf{k}_2)$ is the (localized) detector kernel. It leads to the traced contraction of the second-order moment $\mathcal{M}_a$ with the detector kernel
\begin{equation}
\avg{n} = \tr\{\mathcal{M}_a\diamond D\}-\tfrac{1}{2}\tr\{D\} ,
\label{avgn}
\end{equation}
where the second moment is given by
\begin{align}
\mathcal{M}_a(\mathbf{k}_1,\mathbf{k}_2)
& = \int \alpha(\mathbf{k}_1)\alpha^*(\mathbf{k}_2) W_{\hat{\rho}}[\alpha]\ \Dcirc[\alpha] \nonumber \\
& = \left. -\delta_{\eta}^*(\mathbf{k}_1)\delta_{\eta}(\mathbf{k}_2) \chi_{\hat{\rho}} \right|_{\eta,\eta^*=0} .
\label{defma}
\end{align}

To investigate the evolution of these moments, we use the evolution equation for the characteristic functional in Eq.~(\ref{karevolm}). Thus we can produce an evolution equation for each of the moments.

\subsection{Moment equations}

The zeroth-order moment represents the normalization of the state. It is obtained by setting the auxiliary field variables to zero, leading to
\begin{equation}
\left. \partial_z \chi(z) \right|_{\eta=0} = 0 .
\end{equation}
It shows that the evolution equation for the state (either in terms of the Wigner functional or in terms of the characteristic functional) is trace preserving.

The first-order moments represent the displacement of the state. The resulting equations are
\begin{equation}
\partial_z \mathcal{M}_1 = -\mathcal{M}_1\diamond V_1 ~~~ \text{and} ~~~
\partial_z \mathcal{M}_1^* = -V_1\diamond\mathcal{M}_1^* ,
\end{equation}
where the first-order moments are defined by
\begin{equation}
\mathcal{M}_1 \eqtri \left. \delta_{\eta}^* \chi\right|_{\eta=0} ~~~ \text{and} ~~~
\mathcal{M}_1^* \eqtri \left. -\delta_{\eta} \chi\right|_{\eta=0} .
\end{equation}
Based on the definition of $V_1$ in Eq.~(\ref{defphi1}), the evolution equations simplify to
\begin{equation}
\partial_z \mathcal{M}_1 = -\tfrac{1}{2}\Lambda k^2\mathcal{M}_1 ,
\end{equation}
and its complex conjugate. The solution at $z=L$ is given by the exponential decay of the initial moment:
\begin{equation}
\mathcal{M}_1(L) = \mathcal{M}_1(0) \exp(-\tfrac{1}{2} \Lambda k^2 L) ,
\end{equation}
where $\mathcal{M}_1(0)$ is the first-order moment of the initial state.

There are different second-order moments, represented as kernels. The evolution equation of the Hermitian moment $\mathcal{M}_a$, defined in Eq.~(\ref{defma}), is given by
\begin{align}
\partial_z \mathcal{M}_a(\mathbf{k}_1,\mathbf{k}_2) = &\ 2\int V_0(\mathbf{k}_1,\mathbf{k},\mathbf{k}',\mathbf{k}_2)
\mathcal{M}_a(\mathbf{k},\mathbf{k}')\ \frac{\dm^2 k \dm^2 k'}{(2\pi)^4} \nonumber \\
& -k^2\Lambda\mathcal{M}_a(\mathbf{k}_1,\mathbf{k}_2) ,
\label{maevol}
\end{align}
where we used Eq.~(\ref{defphi1}). In this case, we need to evaluate the integral with the $V_0$-kernel. To solve this evolution equation, we integrate over $z$ from $0$ to $L$ to get
\begin{align}
\mathcal{M}_a(L) = &\ \mathcal{M}_a(0)+\int_0^L 2V_0(z_1)\stackrel{a}{\diamond\diamond}\mathcal{M}_a(z_1) \nonumber \\
& -k^2\Lambda\mathcal{M}_a(z_1)\ \dm z_1 ,
\end{align}
where $\mathcal{M}_a(0)$ is the Hermitian second-order moment of the initial state, and $a$ above the $\diamond\diamond$-contraction indicates the nature of the contraction, as given by the integral in Eq.~(\ref{maevol}). One can now perform back substitutions to replace the evolving second-order moment by the initial second-order moment, up to a given order. The known initial second-order moment then allows one to evaluate the integrals.

The other second-order moments are defined by
\begin{equation}
\mathcal{M}_b \eqtri \left. \delta_{\eta}^{*2}\chi \right|_{\eta=0} ,
\end{equation}
and its complex conjugate. Their evolution equations are
\begin{align}
\partial_z \mathcal{M}_b(\mathbf{k}_1,\mathbf{k}_2) = &\ - k^2\Lambda\mathcal{M}_b(\mathbf{k}_1,\mathbf{k}_2)
- 2 \int \mathcal{M}_b(\mathbf{k},\mathbf{k}') \nonumber \\
& \times V_0(\mathbf{k}_1,\mathbf{k},\mathbf{k}_2,\mathbf{k}')\ \frac{\dm^2 k \dm^2 k'}{(2\pi)^4} ,
\label{mbevol}
\end{align}
and its complex conjugate. Again, we first need to evaluate the integral with $V_0$ before we can solve the equation. Following the same procedure as with $\mathcal{M}_a$, we first integrate over $z$,
\begin{align}
\mathcal{M}_b(L) = &\ \mathcal{M}_b(0)-\int_0^L 2 V_0(z_1)\stackrel{b}{\diamond\diamond}\mathcal{M}_b(z_1) \nonumber \\
& + k^2\Lambda\mathcal{M}_b(z_1)\ \dm z_1 ,
\end{align}
and then perform the back substitutions. In this case, the $b$ above the $\diamond\diamond$-contraction identifies it as given by the integral in Eq.~(\ref{mbevol}).

\section{Coherent state}

As an example, we consider the case where the initial state is a coherent state. Such a scenario is highly relevant to numerous studies of atmospheric scintillation as a classical process. Here, we consider the same scenario, but in the context of a quantum process. The reason for this change in perspective is based on the understanding that, although the process of scintillation is a classical process, the ensemble averaging process also affects the particle-number degrees of freedom of a state, causing a change in the state's photon statistics. In other words, even if the initial state is a coherent state, the evolving state is not a coherent state anymore. This change in the photon statistics can affect the moments that are observed in measurements of the evolving state.

The characteristic functional of a coherent state is
\begin{equation}
\chi_{\text{coh}}[\eta] = \exp\left(-\tfrac{1}{2}\eta^*\diamond\eta+\eta^*\diamond\zeta-\zeta^*\diamond\eta\right) ,
\end{equation}
where $\zeta$ is the parameter function of the coherent state. The second-order moments of the coherent state serve as the initial moments in the integrated versions of their respective evolution equations. They are
\begin{align}
\begin{split}
\mathcal{M}_a(0) & \eqtri \left. -\delta_{\eta}^*\delta_{\eta}\chi_{\text{coh}} \right|_{\eta=0} = \tfrac{1}{2}\mathbf{1}+\zeta\zeta^* , \\
\mathcal{M}_b(0) & \eqtri \left. \delta_{\eta}^{*2}\chi_{\text{coh}} \right|_{\eta=0} = \zeta\zeta .
\end{split}
\end{align}
The integrated forms of the first-order differential equations, under the assumption of an initial coherent state, are then given by
\begin{align}
\begin{split}
\mathcal{M}_a(L) = &\ \tfrac{1}{2}\mathbf{1}+\zeta\zeta^*+\int_0^L 2V_0(z_1)\stackrel{a}{\diamond\diamond}\mathcal{M}_a(z_1) \\
&\ -k^2\Lambda\mathcal{M}_a(z_1)\ \dm z_1 , \\
\mathcal{M}_b(L) = &\ \zeta^2-\int_0^L 2 V_0(z_1)\stackrel{b}{\diamond\diamond}\mathcal{M}_b(z_1) \\
&\ + k^2\Lambda\mathcal{M}_b(z_1)\ \dm z_1 .
\end{split}
\label{intevol}
\end{align}
We now use these equations to perform repeated back substitutions to obtain solutions for the evolution equations of these moments.

First, we consider $\mathcal{M}_a$. The kernel $V_1$ is related to $V_0$ by $V_0\stackrel{a}{\diamond\diamond}\mathbf{1} = V_1 = \tfrac{1}{2}k^2\Lambda\mathbf{1}$, as shown in Eq.~(\ref{phidefs}). It implies that when the identity is substituted for $\mathcal{M}_a$ in the integral terms on the right-hand side of the equation for $\mathcal{M}_a$, they produce
\begin{equation}
2V_0\stackrel{a}{\diamond\diamond}\mathbf{1}-k^2\Lambda\mathbf{1} = 0 .
\end{equation}
This cancellation happens at all orders. The only terms that contribute to the evolution of the moment are those contracted with $\zeta\zeta^*$.

\begin{widetext}
Repeater back substitutions thus produce an infinite series
\begin{align}
\mathcal{M}_a(L) = &\ \tfrac{1}{2}\mathbf{1}+\zeta\zeta^*
+\int_0^L 2V_0(z_1)\stackrel{a}{\diamond\diamond}\zeta\zeta^*
-k^2\Lambda\zeta\zeta^*\ \dm z_1
+\int_0^L \int_0^{z_1} 4V_0(z_1)\stackrel{a}{\diamond\diamond}V_0(z_2)\stackrel{a}{\diamond\diamond}\zeta\zeta^* \nonumber \\
& -2 k^2\Lambda \left[ V_0(z_1) + V_0(z_2) \right]\stackrel{a}{\diamond\diamond}\zeta\zeta^*
+k^4\Lambda^2\zeta\zeta^*\ \dm z_2\ \dm z_1
+\int_0^L \int_0^{z_1} \int_0^{z_2} 8 V_0(z_1)\stackrel{a}{\diamond\diamond} V_0(z_2)
\stackrel{a}{\diamond\diamond} V_0(z_3)\stackrel{a}{\diamond\diamond}\zeta\zeta^* \nonumber \\
& -4 k^2\Lambda \left[ V_0(z_2)\stackrel{a}{\diamond\diamond} V_0(z_3) + V_0(z_1)\stackrel{a}{\diamond\diamond} V_0(z_3)
+ V_0(z_1)\stackrel{a}{\diamond\diamond}V_0(z_2) \right]\stackrel{a}{\diamond\diamond}\zeta\zeta^* \nonumber \\
& +2 k^4\Lambda^2 \left[ V_0(z_1) + V_0(z_2) + V_0(z_3) \right]\stackrel{a}{\diamond\diamond}\zeta\zeta^*
-k^6\Lambda^3 \zeta\zeta^*\ \dm z_3\ \dm z_2\ \dm z_1 + \cdots .
\label{oplma}
\end{align}

For $\mathcal{M}_b$, the back substitutions produce a similar result. Since the initial moment does not contain an identity, we only have the terms with $\zeta^2$. The result is
\begin{align}
\mathcal{M}_b(L) = &\ \zeta^2-\int_0^L 2V_0(z_1)\stackrel{b}{\diamond\diamond}\zeta^2+k^2\Lambda\zeta^2\ \dm z_1
+\int_0^L \int_0^{z_1} 4V_0(z_1)\stackrel{b}{\diamond\diamond}V_0(z_2)\stackrel{b}{\diamond\diamond}\zeta^2 \nonumber \\
& +2 k^2\Lambda \left[ V_0(z_1) + V_0(z_2) \right]\stackrel{b}{\diamond\diamond}\zeta^2
+k^4\Lambda^2\zeta^2\ \dm z_2\ \dm z_1-\int_0^L \int_0^{z_1} \int_0^{z_2} 8 V_0(z_1)\stackrel{b}{\diamond\diamond} V_0(z_2)
\stackrel{b}{\diamond\diamond} V_0(z_3)\stackrel{b}{\diamond\diamond}\zeta^2 \nonumber \\
& +4 k^2\Lambda \left[ V_0(z_2)\stackrel{b}{\diamond\diamond} V_0(z_3) + V_0(z_1)\stackrel{b}{\diamond\diamond} V_0(z_3)
+ V_0(z_1)\stackrel{b}{\diamond\diamond}V_0(z_2) \right]\stackrel{b}{\diamond\diamond}\zeta^2 \nonumber \\
& +2 k^4\Lambda^2 \left[ V_0(z_1) + V_0(z_2) + V_0(z_3) \right]\stackrel{b}{\diamond\diamond}\zeta^2
+k^6\Lambda^3 \zeta^2\ \dm z_3\ \dm z_2\ \dm z_1 + \cdots .
\label{oplmb}
\end{align}
\end{widetext}

\subsection{Average photon number}

The expressions that are derived above can now be used to obtain expressions for quantities that can be measured. For the average photon number, the quantity is shown in Eq.~(\ref{avgn}). Here, we assume a single-mode detector, for which the detector kernel is represented by $D=MM^*$, where the detector mode $M$ is normalized so that $\tr\{D\}=M^*\diamond M=1$. To represent a pixel in a detector array, the detector mode is also parameterized by the coordinates of the output plane. The average photon-number distribution at the output thus becomes
\begin{equation}
\avg{n(\mathbf{x}_0,L)} = M^*\diamond\mathcal{M}_a(\mathbf{x}_0,L)\diamond M-\tfrac{1}{2} ,
\end{equation}
where $\mathbf{x}_0$ is the location of the detector. Using Eq.~(\ref{oplma}), we obtain an expression for this distribution that reads
\begin{widetext}
\begin{align}
\avg{n} = &\ |\mu|^2 +\int_0^L 2Y_1(z_1)-k^2\Lambda|\mu|^2\ \dm z_1
+\int_0^L \int_0^{z_1} 4Y_2(z_1,z_2) -2 k^2\Lambda \left[ Y_1(z_1) + Y_1(z_2) \right] \nonumber \\
& +k^4\Lambda^2|\mu|^2\ \dm z_2\ \dm z_1 +\int_0^L \int_0^{z_1} \int_0^{z_2} 8 Y_3(z_1,z_2,z_3)
-4 k^2\Lambda \left[ Y_2(z_2,z_3) + Y_2(z_1,z_3) + Y_2(z_1,z_2) \right] \nonumber \\
& +2 k^4\Lambda^2 \left[ Y_1(z_1) + Y_1(z_2) + Y_1(z_3) \right]
 -k^6\Lambda^3 |\mu|^2\ \dm z_3\ \dm z_2\ \dm z_1 + \cdots ,
\label{avgnkoh}
\end{align}
\end{widetext}
where
\begin{align}
\begin{split}
\mu \eqtri &\ M^*\diamond\zeta , \\
Y_1(z_1) \eqtri &\ M M^* \stackrel{a}{\diamond\diamond} V_0(z_1)\stackrel{a}{\diamond\diamond}\zeta\zeta^* , \\
Y_2(z_1,z_2) \eqtri &\ M M^* \stackrel{a}{\diamond\diamond}
V_0(z_1)\stackrel{a}{\diamond\diamond}V_0(z_2)\stackrel{a}{\diamond\diamond}\zeta\zeta^* , \\
Y_3(z_1,z_2,z_3) \eqtri &\ M M^* \stackrel{a}{\diamond\diamond} V_0(z_1)\stackrel{a}{\diamond\diamond} V_0(z_2) \\
& \stackrel{a}{\diamond\diamond} V_0(z_3)\stackrel{a}{\diamond\diamond}\zeta\zeta^* .
\end{split}
\label{ydefs}
\end{align}

The overlap between the detector mode and the parameter function is represented by $\mu$, defined in Eq.~(\ref{ydefs}). We use Gaussian functions for both these functions to alleviate the calculation and because the shape of the detector mode is usually not important. The parameter function is given by
\begin{equation}
\zeta(\mathbf{k}) = \sqrt{2\pi} w_0 \zeta_0 \exp\left(-\tfrac{1}{4} w_0^2 |\mathbf{k}|^2 \right) ,
\label{parfunk}
\end{equation}
where $w_0$ is the width of the function on the spatial domain and $\zeta_0$ is the magnitude ($\zeta_0^2$ is the average number of photons in the state). The detector mode is defined as a normalized mode to ensure that the detector kernel is idempotent. However, it needs to allow us to take the limit where the size goes to zero to represent infinite resolution. The normalized detector mode is given by
\begin{align}
M(\mathbf{k},z;\mathbf{x}_0) = &\ \sqrt{2\pi} w_{\text{d}} \exp\left(-\tfrac{1}{4}w_{\text{d}}^2 |\mathbf{k}|^2
-\im \mathbf{k}\cdot\mathbf{x}_0\right) \nonumber \\
& \times \exp\left(-\frac{\im z}{2k} |\mathbf{k}|^2 \right) ,
\label{detfunk}
\end{align}
where $w_{\text{d}}$ is the detector mode size and $\mathbf{x}_0$ is the location on the detector on the output plane. It also contains a $z$-dependent phase factor that allows a final propagation in case the fixed reference frame kernels are used in the calculation. When we use it in the limit of infinite resolution, an additional factor of $2\pi w_{\text{d}}^2$ is produced. Therefore, in the subsequent calculation, we remove this factor to convert the probability into a probability density. The result for $|\mu|^2$, expressed as a probability density reads
\begin{align}
|\mu|^2 = \frac{2 w_0^2 k^2 \zeta_0^2}{(w_0^4 k^2+4L^2)\pi}
\exp\left(-\frac{2w_0^2 k^2|\mathbf{x}_0|^2}{w_0^4 k^2+4L^2}\right) ,
\label{def2mu}
\end{align}
where we set $z=L$.

The calculations of the other overlaps in Eq.~(\ref{ydefs}) become progressively more complicated and tedious for higher orders. In the Appendix, we demonstrate the calculation for $Y_1$. After integrating $|\mu|^2$ over $z_1$ and multiplying it by $k^2\Lambda$, as shown in the subleading (first-order) term in Eq.~(\ref{avgnkoh}), we see that it cancels the term in $2Y_1$ that contains $\Lambda$. The expression for the first-order term thus reads
\begin{align}
& 2Y_1(z_1)-k^2\Lambda|\mu|^2 \nonumber \\
= &\ - \frac{64 \sqrt{2\pi} \mathcal{K} (L-z_1)^{5/3} k\zeta_0^2}{3\Gamma(\tfrac{2}{3})(w_0^4 k^2+4L^2)^{11/6}}
L_{5/6}\left(\frac{2w_0^2 k^2|\mathbf{x}_0|^2}{w_0^4 k^2+4L^2}\right) \nonumber \\
& \times \exp\left(-\frac{2w_0^2 k^2|\mathbf{x}_0|^2}{w_0^4 k^2+4L^2}\right) ,
\end{align}
where \cite{turbsim}
\begin{equation}
\mathcal{K} \eqtri \tfrac{1}{8} w_0^{11/3} k^3 C_n^2 .
\label{defk}
\end{equation}

In the second-order term, similar cancellations remove all its $\Lambda$-terms. We can conclude that such cancellations happen at all orders, thus removing all terms with $\Lambda$'s from the expression of $\avg{n}$. In each order, only one term remains. For the second order, this term is given by
\begin{widetext}
\begin{align}
& 4Y_2(z_1,z_2) -2 k^2\Lambda \left[ Y_1(z_1) + Y_1(z_2) \right]+k^4\Lambda^2|\mu|^2 \nonumber \\
 = &\ \frac{65536\pi^3 2^{1/3} \mathcal{K}^2 (L-z_1)^{5/3}(L-z_2)^{5/3} \zeta_0^2}
 {135\Gamma(\tfrac{2}{3})^5w_0^2(w_0^4 k^2+4L^2)^{8/3}}
L_{5/3}\left(\frac{2w_0^2 k^2|\mathbf{x}_0|^2}{w_0^4 k^2+4L^2}\right)
\exp\left(-\frac{2w_0^2 k^2|\mathbf{x}_0|^2}{w_0^4 k^2+4L^2}\right) ,
\end{align}
and for the third order, it reads
\begin{align}
& 8 Y_3(z_1,z_2,z_3) -4 k^2\Lambda \left[ Y_2(z_2,z_3) + Y_2(z_1,z_3) + Y_2(z_1,z_2) \right]
 +2 k^4\Lambda^2 \left[ Y_1(z_1) + Y_1(z_2) + Y_1(z_3) \right] -k^6\Lambda^3 \nonumber \\
 = &\ - \frac{524288\pi^{11/2} \sqrt{6} \mathcal{K}^3 (L-z_1)^{5/3}(L-z_2)^{5/3}(L-z_3)^{5/3} \zeta_0^2}
 {75\Gamma(\tfrac{2}{3})^9 k w_0^4(w_0^4 k^2+4L^2)^{7/2}}
L_{5/2}\left(\frac{2w_0^2 k^2|\mathbf{x}_0|^2}{w_0^4 k^2+4L^2}\right)
\exp\left(-\frac{2w_0^2 k^2|\mathbf{x}_0|^2}{w_0^4 k^2+4L^2}\right) .
\end{align}
\end{widetext}

These expression forms a pattern that allows us to represent the expression for the photon-number distribution to all orders. First, we evaluate the $z$-integrations. Then we convert all variables and quantities to dimensionless variables and quantities (apart from a factor of $w_0^{-2}$ to provide the units for the distribution as a density). By converting the Laguerre functions to summations, the individual orders are given by
\begin{align}
R_n = &\ \zeta_0^2
\sum_{m=0}^{\infty} \frac{2(-1)^n\kappa^n\beta^{8n/3}\Gamma(1+m+\tfrac{5}{6}n)}{\pi w_0^2(1+\beta^2)^{1+5n/6}n!(m!)^2} \nonumber \\
& \times \left(\frac{-2u_0^2}{1+\beta^2}\right)^m ,
\label{alord}
\end{align}
where we define
\begin{equation}
\kappa \eqtri \frac{2\pi^2\sqrt{6}}{5\Gamma(\tfrac{2}{3})^3} \mathcal{K} = 3.9 \mathcal{K} ,
\label{defkap}
\end{equation}
in terms of Eq.~(\ref{defk}), $u_0\eqtri |\mathbf{x}_0|/w_0$, and
\begin{equation}
\beta \eqtri \frac{2L}{w_0^2 k} = \frac{L}{z_R} ,
\end{equation}
which is inversely proportional to the Fresnel number ($z_R$ is the Rayleigh range). For $n=0$, we have
\begin{equation}
R_0 = \frac{2\zeta_0^2}{\pi w_0^2(1+\beta^2)} \exp\left(\frac{2u_0^2}{1+\beta^2}\right) \equiv |\mu|^2 .
\end{equation}
The photon-number distribution to all orders is
\begin{equation}
\avg{n} = \sum_{n=0}^{\infty} R_n
\end{equation}

\begin{figure}[ht]
\centerline{\includegraphics{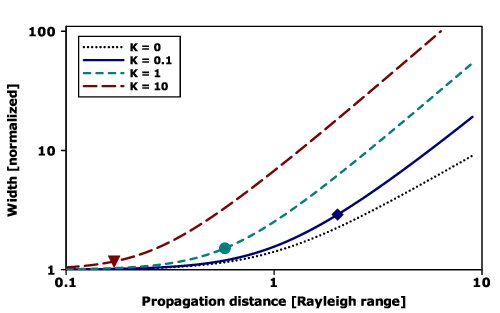}}
\caption{Width of the normalized output photon-number distribution $w(\beta)/w_0$ as a function of the normalized propagation distance $\beta=L/z_R$ for $\mathcal{K}=0,0.1,1,10$, plotted on logarithmic axes. The markers show the points on the respective curves with nonzero $\mathcal{K}$ where the Rytov variance becomes equal to 1 (representing the onset of strong scintillation). }
\label{wydte}
\end{figure}

The evaluation of the summations is difficult due to the factor of $\tfrac{5}{6}$ in the argument of the $\Gamma$-function in Eq.~(\ref{alord}). By approximating $\tfrac{5}{6}\rightarrow 1$ in the argument of the $\Gamma$-function while leaving everything else the same, we can evaluate the summations to obtain
\begin{equation}
\avg{n} \approx \frac{2\zeta_0^2}{\pi(1+\beta^2)(1+\Xi)w_0^2}\exp\left[\frac{-2u_0^2}{(1+\beta^2)(1+\Xi)}\right] ,
\end{equation}
where
\begin{equation}
\Xi \eqtri \frac{\kappa\beta^{8/3}}{(1+\beta^2)^{5/6}} .
\end{equation}
The resulting expression represents a Gaussian profile with a width
\begin{equation}
w(\beta) = w_0(1+\beta^2)^{1/2}\left[1+\frac{\kappa\beta^{8/3}}{(1+\beta^2)^{5/6}}\right]^{1/2} .
\end{equation}
While $\kappa$ represents the strength of the turbulence, $\beta$ represents the propagation distance. In Fig.~\ref{wydte}, we show how the width of the output photon-number distribution increases as a function of the normalized propagation distance $\beta$ for different turbulence strengths. As indications of the onset of strong scintillation, markers are placed at those points along these curves where the Rytov variance, given by
\begin{equation}
\sigma_R^2 = 1.23 C_n^2 k^{7/6} L^{11/6} = 2.76 \mathcal{K} \beta^{11/6} ,
\end{equation}
reaches a value of 1.

\section{Conclusions}

The use of moments in the calculation of measured optical fields in atmospheric scintillation is derived and demonstrated. The moments are obtained from the characteristic functional that can be computed from the Wigner functional for the photonic quantum states of such optical fields. The evolution equation for such Wigner functionals leads to an associated evolution equation for the characteristic functionals, which in turn produces evolution equations for the separate moments. These moments can therefore be solved separately and thus used to compute the desired measured quantities.

To demonstrate this procedure, we consider the scintillation of a coherent state propagating through the atmosphere and compute the observed intensity distribution (photon-number distribution). Its expression is a series expansion in higher orders of the scintillation kernel. Although the calculations of the different orders become progressively more challenging, we compute it to the third order (the details for the calculation of the first order is shown in the Appendix). The results reveal a pattern which allows us to represent the distribution as a summation to all orders. It incorporates the effects of the scintillation process under arbitrary conditions, including those for strong scintillation.

In this calculation, we used the Kolmogorov power spectral density \cite{scintbook}. However, we introduced it in the form of the von Karman spectrum so that the outer scale can be used as a regularization parameter. In the limit of a large outer scale, all the terms associated with this outer scale cancel, leaving only those contributions that are independent of the outer scale.

While our demonstration only considers a quantity that is second order in the field, it is reasonable that the same procedure can be used for quantities that are fourth order in the field, such as the scintillation index. However, it is expected that the associated calculations would become significantly more challenging.

\appendix

\begin{widetext}
\section{First order overlap calculation}

Here, we provide a detailed calculation of $Y_1$, defined in Eq.~(\ref{ydefs}). The kernel given in Eq.~(\ref{defphi0}) is overlapped by four functions, two of which represent the parameter function of the initial coherent state, and the other two represent the detector mode.

To alleviate the calculation, we decouple the wave vectors from the spectral density. The expression for $V_0$ in the fixed reference frame then becomes
\begin{align}
V_0' = &\ 2\pi^2 k^2 \delta(\mathbf{k}_1-\mathbf{k}_2+\mathbf{k}_3-\mathbf{k}_4) \int \Phi_n'(q_0)
\exp\left[\im b_0\left(q_0-|\mathbf{k}_1-\mathbf{k}_2|^2\right)\right]\ \frac{\dm b_0~\dm q_0}{2\pi} \nonumber \\
& \times \exp\left[\frac{\im z}{2k}\left(|\mathbf{k}_1|^2-|\mathbf{k}_2|^2+|\mathbf{k}_3|^2-|\mathbf{k}_4|^2\right)\right] ,
\label{dekphi0}
\end{align}
where we introduce two new integrals. The integration over $b_0$ produces a Dirac delta function that replaces $q_0$ by $|\mathbf{k}_1-\mathbf{k}_2|^2$ in the spectral density. The latter is
\begin{equation}
\Phi_n'(q_0) \eqtri \frac{N_{\text{vK}} C_n^2}{(q_0+\kappa_0^2)^{11/6}} ,
\label{defphin}
\end{equation}
where $N_{\text{vK}}$ is a numerical constant for the von Karman spectral density, $C_n^2$ is the Kolmogorov structure constant and $\kappa_0$ is the inverse of the outer scale of the turbulence.

After the wave vector integrations, we can set $w_{\text{d}}=0$. The result is
\begin{align}
Y_1(z_1) = &\ \int M^*(\mathbf{k}_1,L;\mathbf{x}_0) \zeta(\mathbf{k}_2) \zeta^*(\mathbf{k}_3) M(\mathbf{k}_4,L;\mathbf{x}_0)
V_0'(\mathbf{k}_1,\mathbf{k}_2,\mathbf{k}_3,\mathbf{k}_4,z_1)\
\frac{\dm^2 k_1 \dm^2 k_2 \dm^2 k_3 \dm^2 k_4}{(2\pi)^8} \nonumber \\
= &\ \frac{k^4 w_0^2 \zeta_0^2}{4\pi^3} \exp\left[-\frac{2w_0^2k^2|\mathbf{x}_0|^2}{A_1}\right]
\int \exp\left[\frac{8 A_0^2k^2|\mathbf{x}_0|^2}{(A_2+\im 2 A_1 b_0)A_1}\right]
\frac{\exp(\im b_0 q_0) \Phi_n'(q_0)}{A_2+\im 2 A_1 b_0}\ \dm b_0~\dm q_0 ,
\label{ey1}
\end{align}
where
\begin{equation}
A_0 \eqtri (L-z_1)w_0^2 , ~~~~~
A_1 \eqtri w_0^2 k^2+4 L^2 , ~~~~~ \text{and} ~~~~~
A_2 \eqtri 4 w_0^2 (L-z_1)^2 .
\label{aadefs}
\end{equation}

We use contour integration to evaluate the integral over $b_0$. The integrand has a pole at
\begin{equation}
b_0 = \frac{\im A_2}{2 A_1} \eqtri \im b_1 ,
\end{equation}
in the upper half plane, because $A_2/2 A_1>0$. The integral along the arch at infinity is zero provided that $q_0>0$. The contour integral is then of the form
\begin{equation}
\oint \frac{\exp(\im b_0 q_0)}{b_0-\im b_1} \exp\left(\frac{-\im c_0}{b_0-\im b_1}\right)\ \dm b_0
= \sum_m \oint \frac{(-\im c_0)^m \exp(\im b_0 q_0)}{(b_0-\im b_1)^{m+1} m!}\ \dm b_0
= \im 2\pi \sum_m \frac{ (q_0 c_0)^m \exp(- b_1 q_0)}{m!^2} ,
\end{equation}
where
\begin{equation}
c_0 \eqtri \frac{4 A_0^2 k^2|\mathbf{x}_0|^2}{A_1^2} .
\end{equation}
Combined with the rest of the expression, it becomes
\begin{align}
Y_1(z_1) = &\ \frac{k^4 w_0^2\zeta_0^2}{4\pi^2 A_1} \exp\left[-\frac{2w_0^2k^2|\mathbf{x}_0|^2}{A_1}\right]
\int \sum_m \frac{1}{m!^2} \left(\frac{4 A_0^2 k^2|\mathbf{x}_0|^2 q_0}{A_1^2}\right)^m
\exp\left(-\frac{A_2 q_0}{2 A_1}\right) \Phi_n'(q_0)\ \dm q_0 .
\end{align}
For a large enough outer scale, the integration over $q_0$ produces
\begin{align}
Q_m & = \int q_0^m \exp\left(-\frac{A_2 q_0}{2 A_1}\right) \Phi_n'(q_0)\ \frac{\dm q_0}{2\pi}
= 4\pi\Lambda \delta_{0,m} - \frac{2\pi(-1)^m N_{\text{vK}} C_n^2}{\Gamma(\tfrac{11}{6}-m)} \left(\frac{A_2}{2 A_1}\right)^{5/6-m} ,
\end{align}
with $\Lambda$ defined in Eq.~(\ref{deflam}). We combine it with the rest of the expression and evaluate the summation to obtain
\begin{align}
Y_1(z_1) = &\ \frac{w_0^2 k^4 \zeta_0^2 \Lambda}{(w_0^4 k^2+4z^2)\pi}
\exp\left(-\frac{2w_0^2 k^2|\mathbf{x}_0|^2}{w_0^4 k^2+4z^2}\right) \nonumber \\
& - \frac{32 k 2^{2/3} (L-z_1)^{5/3} \zeta_0^2 \kappa}{3(w_0^4 k^2+4z^2)^{11/6}\pi}
L_{5/6}\left(\frac{2w_0^2 k^2|\mathbf{x}_0|^2}{w_0^4 k^2+4z^2}\right)
\exp\left(-\frac{2w_0^2 k^2|\mathbf{x}_0|^2}{w_0^4 k^2+4z^2}\right) ,
\end{align}
where $L_{\nu}(\cdot)$ is the Laguerre function and $\kappa$ is defined in  Eq.~(\ref{defkap}).
\end{widetext}


\end{document}